# Iodide-methylammonium interaction is responsible for ferroelectricity in $CH_3NH_3PbI_3$

**Authors:** J. Breternitz[1]*, F. Lehmann[1,2], S. A. Barnett[3], H. Nowell[3], S. Schorr[1,4].

**Affiliations:**

[1]Struktur und Dynamik von Energiematerialien, Helmholtz-Zentrum Berlin für Materialien und Energie, Hahn-Meitner Platz 1, 14109 Berlin, Germany.

[2]Insititute of Chemistry, Universität Postdam, 14469 Potsdam, Germany.

[3]Diamond Light Source, Didcot OX11 0DE, United Kingdom.

[4]Department of Geosciences, Freie Universität Berlin, Malteserstraße 74-100, 12249 Berlin, Germany.

*Correspondence to: Joachim.breternitz@helmholtz-berlin.de

**Abstract:** Excellent conversion efficiencies of over 20 % and facile cell production have placed hybrid perovskites at the forefront of novel solar cell materials with $CH_3NH_3PbI_3$ being its archetypal compound. The question why $CH_3NH_3PbI_3$ has such extraordinary characteristics, particularly a hugely efficient light absorption, is hotly debated with ferroelectricity being a promising candidate. This does, however, afford the crystal structure to be non-centrosymmetric and we herein present crystallographic evidence as to how the symmetry breaking occurs on a crystallographic, and therefore long-scale, level. While the molecular cation $CH_3NH_3^+$ is intrinsically polar, it is heavily disordered and cannot be the sole reason for ferroelectricity. We show that it, nonetheless, plays an important role as it distorts the neighboring iodide positions from their centrosymmetric positions.

**Main Text:** It is undoubtable that hybrid perovskites have changed the way we are looking at solar absorber materials (1-4). Traditionally, semiconductors were thought as rigid solids with highly defined atom positions. Hybrid perovskites, however, were shown to have a high defect tolerance (5) and a flexible crystal structure with remarkable positional freedom of the molecular cation (6) and ionic movement (7,8). This latter makes a reliable crystal structure determination challenging as the average long range order no longer reflects all the properties of the material. It is probably also due to this fact that no real consensus was reached as to whether $CH_3NH_3PbI_3$ at room temperature is centrosymmetric or not (9). While many bulk and thin film measurements indicate a ferroelectric effect of $CH_3NH_3PbI_3$ at ambient conditions (10-13), other studies either could not reproduce this effect or come to a different conclusion (14-17). Besides the direct observation of ferroelectric response, a crystallographic prerequisite exists for ferroelectricity: the crystal structure must be polar, *i.e.* belong to a space group that is not only non-centrosymmetric, but must belong to one of the 10 polar crystal classes (18). The common crystal structure of $CH_3NH_3PbI_3$ at room temperature, however, is given in space group *I*4/*mcm* (9,19), which is centrosymmetric and hence would not allow any of the above mentioned effects. Herein, we set out to conduct high resolution single crystal diffraction to elaborate the reason for the observed polarizability of $CH_3NH_3PbI_3$ combined with a discussion of the possible space group setting of the compound.

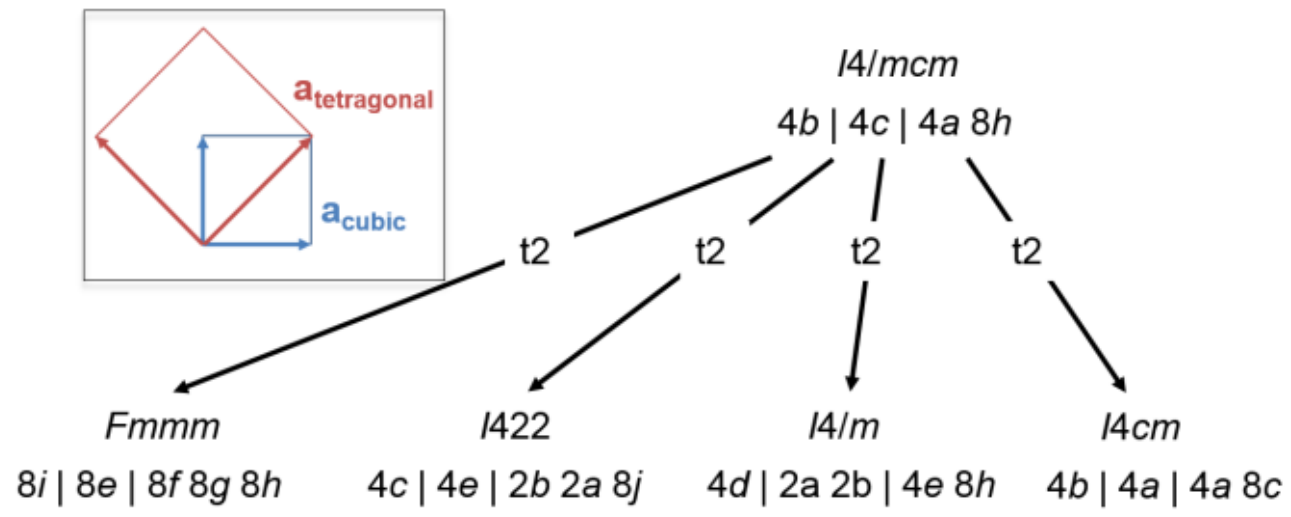

**Fig. 1.** group-subgroup relationships between the common space group *I*4/*mcm* and further space groups found in literature. Inset: relation of the lattice vectors in the *ab*-plane between the cubic aristotype and the tetragonal setting.

**Space group of $CH_3NH_3PbI_3$**

While *I*4/*mcm* is the commonly chosen space group, numerous other choices are documented in the literature (figure 1). These choices roughly fall in two categories: 1) space groups that no longer contain the *c*-glide plane (20-22) and 2) the space group *I*4*cm* (23), which is the only polar maximal subgroup of *I*4/*mcm*. It is possible to refine the crystal structure in any of the given space groups: since all alternative choices are subgroups of the common choice *I*4/*mcm*, they all contain a subset of symmetry elements, but no symmetry elements that would not exist in *I*4/*mcm*. Therefore, a crystal structure in *I*4/*mcm* must also contain all symmetry elements of the lower symmetry hettotypes. When comparing the atomic parameters, however, between the different refinements, it becomes evident that all refined structures are closely related to each other.

In fact, the space groups falling in 1) were chosen by the authors because they observed supplementary reflections, which violate the systematic extinctions dictated by the translational symmetry element, the *c*-glide plane (20). However, this apparent symmetry breaking is most probably due to twinning of the single crystals in these studies and is aggravated through the nature of the material: the cubic-to-tetragonal transition in the system is relatively close to room temperature, signifying that the energy difference between the two at ambient conditions is marginal. Therefore, one could easily assume that the crystal nucleation points form in the cubic symmetry and only the bulk material is tetragonal. If this was true, the choice of the *c*-axis out of the three equivalent axes in the cubic system is arbitrary and could easily change within a crystal and one would expect axis twins. While $CH_3NH_3PbI_3$ at room temperature is tetragonal, the mismatch between the crystallographic *c-axis* on the one hand and the crystallographic *a*- and *b*-axes (with $a$=$b$) on the other hand is below 1 %. This is not directly visible when looking at the lattice constants, because the tetragonal lattice constants are related to the cubic ones through $a_{\text{tetragonal}} = \sqrt{2} \cdot a_{\text{cubic}}$ (including a 45° shift, see figure 1 inset) and $c_{\text{tetragonal}} = 2 \cdot c_{\text{cubic}}$. Axis twinning would therefore not necessarily result in extensive peak splitting or supplementary reflections, as the twinned reflections almost perfectly overlap with the main reflections, apart from the positions where the systematic extinctions of the main reflections should lie. Including the appropriate twin law in the refinement of the data provided by the original authors suppresses the systematic extinction violations entirely and hence supports the explanation of the apparent extinction violation through twinning effects (See SI for detailed analysis). Further, the splitting of the Pb or I position induced by the symmetry descent is not reflected in the atomic positions. In fact, since the space groups under discussion possess a different translational symmetry to *I*4/*mcm*, they should also show supplementary reflections in powder diffraction (24), but no such supplementary reflection was documented.

The situation for 2) is different: *I4cm* does not add additional splitting in the atomic positions, but allows more positional freedom for the atomic positions. The most striking difference between *I*4/*mcm* and *I*4*cm* is the lack of mirror plane perpendicular to the four-fold axis, *i.e.* in the ab-plane. This allows the atomic positions to move arbitrarily along the crystallographic *c*-axis and hence allows a shift of the atoms outside a common plane. Such a shift induces a permanent polar moment and hence can induce ferroelectricity. While the molecular cation $CH_3NH_3^+$ is intrinsically polar, it is dynamically and statically disordered (25) and therefore probably does not induce an effective macroscopic moment. In order to explain this evident mismatch with the experimental evidence for ferroelectricity, we performed high-resolution synchrotron single crystal diffraction to study the atomic positions at the best accuracy possible.

**Molecular cation orientation and iodine shifting**

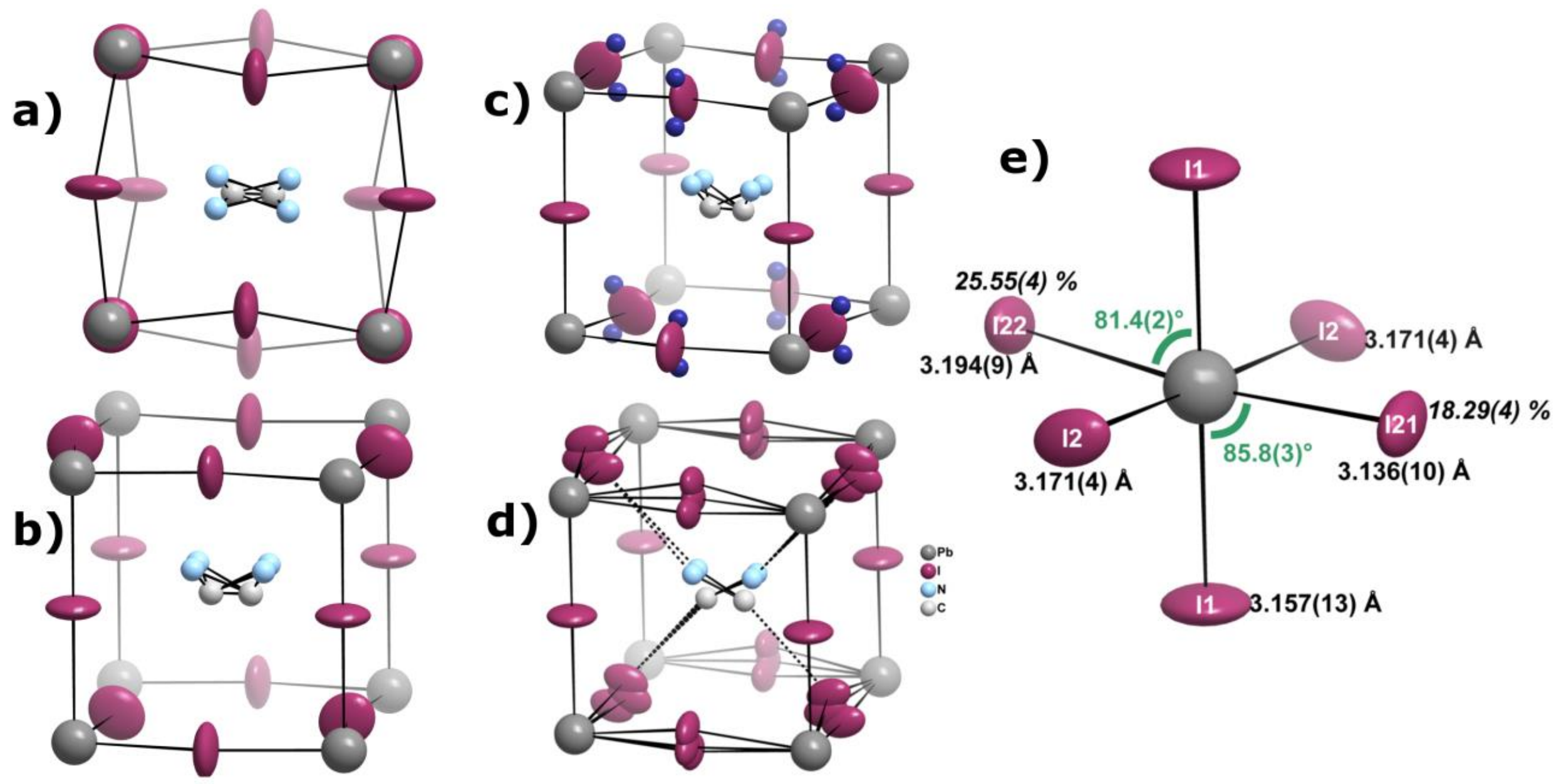

**Fig. 2.** Structural peculiarities of $MAPbI_3$ at room temperature. Orientation of the $CH_3NH_3^+$ cation in the pseudo-cubic [$PbI_3$] cage along the c-axis (a) and in a general section (b) in a conventional 2 I-site refinement. Illustration of the highest residual electron density peaks in the 2 I-site model (dark blue dots, c) and the cage including the split-iodine positions (d). Representation of the $PbI_6$ octahedron including the split sites. Pb-I distances are given in black, the relative occupancies of I21 and I22 in italic and the I1-Pb-I21/I22 angles in green.

Conventionally, ferroelectric perovskites are showing a shift of the cations (26). This shift can be very small indeed, as was recently shown in the ferroelectric phase of $SrTiO_3$ (27). Therefore, we performed single crystal diffraction at the Pb L-absorption edges. Under these conditions, the complex part of the atomic structure factor is maximal and can become non-negligible. Therefore, Friedel's law is not strictly valid any more. Briefly: as the intensity of reflections is proportional to the atomic structure factors, it will be equal for hkl and $\overline{h}\overline{k}\overline{l}$ reflections. When the complex part of the structure factor becomes non-negligible, this is no longer true and the observation of such Bijvoet pairs would be a direct proof of a non-centrosymmetric structure. However, we have not found a breaking of Friedel's law, which does not directly rule the possibility of non-centrosymmetricity out, but is most likely an effect of inversion twinning in the crystals under consideration. In fact, several recent studies have observed twin domains in $MAPbI_3$ crystals, both in thin films and bulk crystals (28-30).

Using the space group *I*4*cm* instead of *I*4/*mcm* during the crystal structure refinement did yield a refinement that is very similar to the structures reported in the literature, without a clear shift of the relevant atoms (see SI). However, two points are distinctly different to the refinements in *I*4/*mcm*: 1) the orientation of the molecular cation is less disordered than in previous studies (9,19,31) and only shows two distinct orientations (figure 2 a and b) and 2) the highest residual electron density peaks (+3.24 $e^-/Å^3$) in the system are close to I2, which is the iodine site within the *ab*-plane. The differences between the molecular cation orientation in this refinement and previous studies is only seemingly contradictive: the supplementary positions found in those previous studies aiming at elucidation the molecular orientation are a direct consequence of the higher symmetry in *I*4/*mcm*. These studies were based on powder diffraction, but the differences between *I*4/*mcm* and *I*4*cm* are invisible in powder diffraction as the reflections being different perfectly overlap with each over. 2) Normally, one would expect the highest residual electron density peaks close to the heaviest atoms, Pb in here, as an effect of the finite number of elements in the Fourier transformation series. Having them at the iodine position instead is probably due to missing the assignment of some electron density. Indeed, this can be easily interpreted as partial occupation of iodine distributed over three atomic sites, of which two are outside the *ab*-plane. It should be noted that similar residual electron density peaks can be found in the datasets by Jaffe et al. (20) and Arakcheeva et al. (21).

Both those effects, discussed in 1) and 2) above, are in fact related to each other and probably influence each other. It is clear that the molecular cation is roughly pointing towards two of the iodine atoms at opposite edges of the roughly cube-shaped cage (figure 2). This is easily understandable as this orientation maximizes the hydrogen bonding interaction between the molecular cation and the surrounding iodine atoms. On the other hand, the iodine atoms positioned out of the plane are shifted in a way that they specifically approach the molecular cations at the edges the cations point at, while they are shifted away at the other edges (figure 2 d). This is in line with the general argument of maximizing X-H···I (X = C, N) interactions as it allows shorter H-I distances. Further, the shifting is observed above and below the *ab*-plane, but the vector between the two shifted positions is not perpendicular to the *ab*-plane but stands at an angle of 62.5° to the plane. This is tremendously important, as a perpendicular distortion would still be explainable in *I*4/*mcm*, while such a shifted situation clearly is not and hence explains the non-centrosymmetric arrangement of $CH_3NH_3PbI_3$ at room temperature.

Clearly, the arrangement of the inorganic framework and the arrangement of the molecular cation influence each other in this compound and one might ask the question as to whether the molecular arrangement causes the iodine shift or whether the iodine shift causes a locking of the molecular cation. The consequences of the iodine shifting on the arrangement in the [$PbI_6$] octahedra (figure 2e) is not great bust distinctive: the Pb-I distances become more anisotropic but the differences are generally below 0.1 Å and the I-Pb-I angles are slightly distorted from the ideal 90° arrangement. Using diffraction techniques, it is impossible to answer this question directly because of the time and space averaging in diffraction. However, the great strength of this explanation is that it does not solely rely on the intrinsic polarity of the molecular cation, which is both statically and dynamically disordered, but its relation with the surrounding iodine atoms. In fact, no matter how the individual molecular cation is ordered in each individual cage, its relationship with the iodine atoms will be similar so that the shift of the iodine atoms remain relatively constant creating macroscopic polarity of the compound.

**Conclusion**

This finding is of crucial importance for the understanding of hybrid perovskites: it not only makes the finding of ferroelectric effects in $CH_3NH_3PbI_3$ at ambient conditions reasonable but it effectively explains where it comes from: the interaction of the molecular cation with the anion framework. Therefore, the unique properties of the hybrid perovskites do critically depend on the nature of the organic cation and it will be important, in a subsequent step, to assess how this changes when modifying the molecular cation. This in fact also points to why all-inorganic perovskites do not exhibit the same efficiencies as hybrid perovskites do. This finding imminently raises the fundamental question whether the desired effects of the molecular cation, most prominently high efficiency, can be preserved while targeting its negative sides, especially the operation stability under light, or whether this poses a critical intrinsic dilemma of these compounds that cannot be overcome.

**Acknowledgments:** The authors would like to acknowledge Natalie Johnson and Mike Probert for using their data conversion program. We are deeply grateful to Prof. Hartmut Bärnighausen for discussion on the topic and Dr. David Allan for discussion on the beamtime proposal. **Funding:** The authors acknowledge Diamond Light Source for the provision of beamtime under award number MT17170 and the HyperCells graduate school for funding the PhD of FL; **Author contributions:** JB and SuS designed the study and SuS supervised the work, JB, FL, HN and SAB conducted the experiments, JB undertook the data processing and wrote the initial manuscript, all authors contributed in proof-reading and gave consent to the final version; **Competing interests:** Authors declare no competing interests.

## Supporting information

## Iodide-methylammonium interaction is responsible for ferroelectricity in $CH_3NH_3PbI_3$

Joachim Breternitz, Frederike Lehmann, Sarah A. Barnett, Harriot Nowell, Susan Schorr

## 1. Twinning in Arakcheeva et al. (21) and Jaffe et al. (20)

To test for possible twinning as cause for the supposedly observed breaking of the c-glide plane, a twin-law according to pseudo-merohedral pseudo-cubic axial twinning using the command:

```
TWIN 0.5 0.5 -0.5 0.5 0.5 0.5 1 -1 0
```

According to a 90° rotation around the <110> direction in the tetragonal unit cell.

**SHELXL list file output for the untwinned model in Arakcheeva et al. (21) using *I4cm*:**

```
   h   k   l        Fo^2      Sigma    Why rejected (first 50 of each listed)
   0   5   7        0.08      0.01     systematically absent but >3sig(I)

       35  Systematically absent reflections rejected


      773  Reflections read, of which      35  rejected

  -8 =< h =<  9,      0 =< k =< 13,      0 =< l =< 13,   Max. 2-theta =   62.53

        1  Systematic absence violations (I>3sig(I)) before merging

       13  Inconsistent equivalents

      432  Unique reflections, of which       0  suppressed

 R(int) = 0.0083     R(sigma) = 0.0050      Friedel opposites not merged

 Maximum memory for data reduction =      987 /     5340


 Number of data for d > 0.659A (CIF: max) and d > 0.833A (CIF: full)
 (ignoring systematic absences):
 Unique reflections found (point group)        432     254
 Unique reflections possible (point group)    1004     510
 Unique reflections found (Laue group)         432     254
 Unique reflections possible (Laue group)      524     269
 Unique Friedel pairs found                      0       0
 Unique Friedel pairs possible                 480     241
```

**SHELXL list file output for the twinned model in Arakcheeva et al. using *I4cm* (Approximate twin fraction 1%):**

```
      773  Reflections read, of which       0  rejected

  -8 =< h =<  9,      0 =< k =< 13,      0 =< l =< 13,   Max. 2-theta =   62.53
```

```
       0  Systematic absence violations (I>3sig(I)) before merging

      13  Inconsistent equivalents

     462  Unique reflections, of which       0  suppressed

 R(int) = 0.0083     R(sigma) = 0.0056      Friedel opposites not merged

 Maximum memory for data reduction =      988 /     5620


 Number of data for d > 0.659A (CIF: max) and d > 0.833A (CIF: full)
 (ignoring systematic absences):
 Unique reflections found (point group)        748     413
 Unique reflections possible (point group)    1004     510
 Unique reflections found (Laue group)         443     256
 Unique reflections possible (Laue group)      524     269
 Unique Friedel pairs found                    305     157
 Unique Friedel pairs possible                 480     241
```

The raw hkl data from Jaffe *et al.* was brought in the right setting for *I4cm* using Jana2006 (32).

**SHELXL list file output for the untwinned model in Jaffe et al. (20) using *I4cm*:**

```
   h   k   l        Fo^2      Sigma    Why rejected (first 50 of each listed)
   0   3  -1        6.96      0.50     systematically absent but >3sig(I)
   0   3   1        7.72      0.44     systematically absent but >3sig(I)
   0   3  -1        8.62      0.50     systematically absent but >3sig(I)
   0   3   1        7.66      0.38     systematically absent but >3sig(I)
   0   3   1        7.85      0.40     systematically absent but >3sig(I)
   0   3  -1        7.00      0.45     systematically absent but >3sig(I)
   0   3   1        7.43      0.42     systematically absent but >3sig(I)
   0   3   1        7.91      0.52     systematically absent but >3sig(I)
   0   3  -1        8.21      0.44     systematically absent but >3sig(I)
   0   3  -1        7.51      0.36     systematically absent but >3sig(I)
   0   3  -1        7.72      0.39     systematically absent but >3sig(I)
   0   3   1        8.22      0.48     systematically absent but >3sig(I)
   0   3   1        7.21      0.38     systematically absent but >3sig(I)
   0   5  -1        1.24      0.25     systematically absent but >3sig(I)
   0   5   1        0.62      0.17     systematically absent but >3sig(I)
   0   5   1        0.71      0.20     systematically absent but >3sig(I)
   0   5  -1        1.12      0.28     systematically absent but >3sig(I)
   0   5   1        0.96      0.18     systematically absent but >3sig(I)
   0   5   1        0.60      0.13     systematically absent but >3sig(I)
   0   5  -1        0.52      0.16     systematically absent but >3sig(I)
   0   5   1        0.88      0.22     systematically absent but >3sig(I)
   0   5   1        0.90      0.19     systematically absent but >3sig(I)
   0   5  -1        0.98      0.15     systematically absent but >3sig(I)
   0   5   1        0.80      0.24     systematically absent but >3sig(I)
   0   5  -1        0.95      0.17     systematically absent but >3sig(I)
   0   5  -1        0.84      0.24     systematically absent but >3sig(I)
   0   7   1        0.57      0.18     systematically absent but >3sig(I)
   0   7   1        0.83      0.23     systematically absent but >3sig(I)
   0   1  -3        7.87      0.41     systematically absent but >3sig(I)
   0   1   3        8.67      0.44     systematically absent but >3sig(I)
   0   1  -3        7.42      0.52     systematically absent but >3sig(I)
   0   1  -3        8.43      0.39     systematically absent but >3sig(I)
   0   1   3        9.29      0.44     systematically absent but >3sig(I)
   0   1   3        8.76      0.40     systematically absent but >3sig(I)
```

```
  0   1   3        8.12      0.56     systematically absent but >3sig(I)
  0   5  -3        6.82      0.59     systematically absent but >3sig(I)
  0   5  -3        5.59      0.50     systematically absent but >3sig(I)
  0   5   3        6.75      0.49     systematically absent but >3sig(I)
  0   5  -3        4.72      0.49     systematically absent but >3sig(I)
  0   5   3        6.84      0.41     systematically absent but >3sig(I)
  0   5  -3        7.62      0.53     systematically absent but >3sig(I)
  0   5  -3        6.79      0.49     systematically absent but >3sig(I)
  0   5   3        5.70      0.47     systematically absent but >3sig(I)
  0   1   5        0.44      0.14     systematically absent but >3sig(I)
  0   1  -5        0.82      0.22     systematically absent but >3sig(I)
  0   1   5        0.42      0.09     systematically absent but >3sig(I)
  0   1   5        0.44      0.13     systematically absent but >3sig(I)
  0   1   5        0.31      0.10     systematically absent but >3sig(I)
  0   3  -5       10.01      0.71     systematically absent but >3sig(I)
  0   3  -5       10.92      0.57     systematically absent but >3sig(I)

** etc. **

      393  Systematically absent reflections rejected


     7624  Reflections read, of which     393  rejected

-13 =< h =< 14,    -14 =< k =< 14,    -20 =< l =< 20,   Max. 2-theta =   70.06

       76  Systematic absence violations (I>3sig(I)) before merging

       21  Inconsistent equivalents

     1280  Unique reflections, of which       0  suppressed

R(int) = 0.0628     R(sigma) = 0.0375      Friedel opposites not merged

Maximum memory for data reduction =      975 /    16887


Number of data for d > 0.600A (CIF: max) and d > 0.833A (CIF: full)
(ignoring systematic absences):
Unique reflections found (point group)       1280     494
Unique reflections possible (point group)    1291     496
Unique reflections found (Laue group)         669     262
Unique reflections possible (Laue group)      671     262
Unique Friedel pairs found                    611     232
Unique Friedel pairs possible                 620     234
```

**SHELXL list file output for the twinned model in Jaffe et al. using *I4cm* (Approximate twin fraction 12%):**

```
     7624  Reflections read, of which       0  rejected

-13 =< h =< 14,    -14 =< k =< 14,    -20 =< l =< 20,   Max. 2-theta =   70.06

        0  Systematic absence violations (I>3sig(I)) before merging

       21  Inconsistent equivalents

     1385  Unique reflections, of which       0  suppressed

R(int) = 0.0633     R(sigma) = 0.0385      Friedel opposites not merged

Maximum memory for data reduction =      988 /    17823
```

```
Number of data for d > 0.596A (CIF: max) and d > 0.833A (CIF: full)
(ignoring systematic absences):
Unique reflections found (point group)       1284     494
Unique reflections possible (point group)    1316     496
Unique reflections found (Laue group)         673     262
Unique reflections possible (Laue group)      684     262
Unique Friedel pairs found                    611     232
Unique Friedel pairs possible                 632     234
```

## 2. Experimental details

Crystals were grown at room temperature according to the antisolvent vapor method described by Rakita et al. $PbI_2$ (99 %, ACROS Organics), ethyl acetate (ChemCruz, HPLC grade), acetonitrile (Sigma-Aldrich, 99.5 %), diethyl ether (Merck, 99.7 %), methylammonium iodide (Sigma-Aldrich, 98 %) and HI solution (stabilized 57 wt.-%in $H_2O$, 99.95 %, Sigma-Aldrich) were used as supplied. Both, crystals grown with diethyl ether and ethyl acetate as antisolvent were tested, but those grown using ethyl acetate generally exhibited better crystal quality and the study was conducted on a crystal of this series. It should be noted that we did not find a single crystal which did not show any signs of twinning and finally selected one that appeared least twinned for the subsequent detailed analysis. Crystals were prepared in an Ar-filled glovebox and covered in oil during the measurement to avoid sample decomposition due to moisture. It should be emphasized that crystals with approx. edge lengths of 20 µm were used for these experiments to avoid further complications with heavy twinning and strong absorption.

Single-crystal X-ray diffraction was conducted at the I19 beamline at the Diamond Light Source synchrotron. Using the double crystal monochromator of the beamline, the X-ray energy was adjusted between 12.97 keV and 15.3 keV, i.e. in proximity of the L-III and L-II absorption edges of lead. This was chosen as the initial approach of this experiment was to test, whether a breaking of Friedel's law as direct proof for the lack of inversion symmetry could be observed. It should be emphasized that we did not observe any significant breaking of Friedel's law, which is most probably due to the domain nature of the crystals. In fact, we refined the final model as inversion twin yielding in a twin fraction of 48 %. In order to observe Friedel pairs, one would probably need to align the domains, for instance through crystallization in an electric field. We are currently testing such possibilities. Given no direct observation of Friedel pairs could be achieved, further analysis was performed on the highest measured energy: 15.3 keV ($\lambda$ = 0.81036 Å).

Reflections were measured using a Pilatus 2M detector. Data integration and Lorentz factor correction (using SAINT V8.38A) and absorption correction (using SADABS-2016/2) were performed using the Bruker APEX3 suite (33), for which the Pilatus CBF format was converted to SFRM using a custom built program by Natalie Johnson and Mike Probert (34). The authors are thankful for their kind help with this. The latter was done using a semi-empirical multiscan absorption correction as the crystal form could not be reliably determined given the size and the covering in oil. Refinements were performed using SHELXL2013 (35).

The C-N distance of the $CH_3NH_3^+$ cation was fixed to 1.47 Å as common for the molecular cation (19). The split iodine sites in the split-site model were constrained to have equal displacement parameters. This refinement was further damped at the later stages since the molecular cation is heavily disordered. It should be noted that the assignment of carbon and nitrogen in the model is arbitrary, since the small difference in electron density between carbon and nitrogen makes them literally indistinguishable, particularly in connection with iodine and lead.

## 3. Refinement without split site model

*Table S1: Crystal data*

| | |
|---|---|
| $CH_6I_3NPb$ | $Z = 4$ |
| $M_r = 619.96$ | $F(000) = 1040$ |
| Tetragonal, *I4cm* | $D_x = 4.171$ Mg m$^{-3}$ |
| $a = 8.8438$ (3) Å | $\mu = 26.39$ mm$^{-1}$ |
| $c = 12.6215$ (5) Å | $T = 293$ K |
| $V = 987.16$ (8) Å$^3$ | |

*Table S2: Data collection*

| | |
|---|---|
| 5668 measured reflections | $\theta_{max} = 34.5°$, $\theta_{min} = 3.7°$ |
| 768 independent reflections | $h = -12 \rightarrow 12$ |
| 674 reflections with $I > 2\sigma(I)$ | $k = -12 \rightarrow 12$ |
| $R_{int} = 0.054$ | $l = -17 \rightarrow 17$ |

*Table S3: Refinement*

| | |
|---|---|
| Refinement on $F^2$ | H-atom parameters not defined |
| Least-squares matrix: full | $w = 1/[\sigma^2(F_o^2) + (0.0569P)^2 + 22.4079P]$ where $P = (F_o^2 + 2F_c^2)/3$ |
| $R[F^2 > 2\sigma(F^2)] = 0.044$ | $(\Delta/\sigma)_{max} < 0.001$ |
| $wR(F^2) = 0.126$ | $\Delta\rangle_{max} = 3.24$ e Å$^{-3}$ |
| $S = 1.15$ | $\Delta\rangle_{min} = -1.34$ e Å$^{-3}$ |
| 768 reflections | Absolute structure: Flack x determined using 291 quotients [(I+)-(I-)]/[(I+)+(I-)] (Parsons, Flack and Wagner, Acta Cryst. B69 (2013) 249-259). |
| 20 parameters | Absolute structure parameter: 0.48 (3) |
| 2 restraints | |

*Table S4: Fractional atomic coordinates and isotropic or equivalent isotropic displacement parameters (Å$^2$)*

| | *x* | *y* | *z* | $U_{iso}*/U_{eq}$ | Occ. (<1) |
|---|---|---|---|---|---|
| Pb | 0.000000 | 0.000000 | 0.00012 (2) | 0.0298 (3) | |
| I1 | 0.000000 | 0.000000 | 0.2500 (4) | 0.0812 (10) | |
| I2 | 0.2135 (2) | 0.7135 (2) | 0.0007 (7) | 0.0844 (8) | |
| N | 0.404 (7) | 0.036 (8) | 0.278 (6) | 0.05 (2)* | 0.25 |
| C | 0.537 (4) | -0.037 (4) | 0.229 (5) | 0.048 (15)* | 0.5 |

*Table S5: Atomic displacement parameters (Å$^2$)*

| | $U^{11}$ | $U^{22}$ | $U^{33}$ | $U^{12}$ | $U^{13}$ | $U^{23}$ |
|---|---|---|---|---|---|---|
| Pb | 0.0311 (4) | 0.0311 (4) | 0.0273 (5) | 0.000 | 0.000 | 0.000 |
| I1 | 0.1123 (16) | 0.1123 (16) | 0.0190 (9) | 0.000 | 0.000 | 0.000 |
| I2 | 0.0705 (9) | 0.0705 (9) | 0.1121 (17) | 0.0480 (10) | 0.010 (3) | 0.010 (3) |

*Table S6: Geometric parameters (Å, º)*

| | | | |
|---|---|---|---|
| Pb—I1 | 3.154 (5) | N—C$^{vi}$ | 0.81 (7) |
| Pb—I1$^{i}$ | 3.157 (5) | N—N$^{vii}$ | 0.74 (14) |
| Pb—I2$^{ii}$ | 3.1600 (4) | N—C | 1.47 (3) |
| Pb—I2$^{iii}$ | 3.1600 (4) | N—N$^{viii}$ | 1.65 (12) |
| Pb—I2$^{iv}$ | 3.1600 (4) | C—C$^{vi}$ | 0.92 (10) |
| Pb—I2$^{v}$ | 3.1600 (4) | | |
| | | | |
| I1—Pb—I1$^{i}$ | 180.0 | C$^{vi}$—N—N$^{vii}$ | 63 (5) |
| I1—Pb—I2$^{ii}$ | 89.87 (16) | C$^{vi}$—N—C | 34 (6) |
| I1$^{i}$—Pb—I2$^{ii}$ | 90.13 (16) | N$^{vii}$—N—C | 75 (3) |
| I1—Pb—I2$^{iii}$ | 89.87 (16) | C$^{vi}$—N—N$^{viii}$ | 63 (7) |
| I1$^{i}$—Pb—I2$^{iii}$ | 90.13 (16) | N$^{vii}$—N—N$^{viii}$ | 90.002 (11) |
| I2$^{ii}$—Pb—I2$^{iii}$ | 179.7 (3) | C—N—N$^{viii}$ | 29 (3) |
| I1—Pb—I2$^{iv}$ | 89.87 (16) | C$^{vi}$—C—N$^{vi}$ | 117 (7) |
| I1$^{i}$—Pb—I2$^{iv}$ | 90.13 (16) | C$^{vi}$—C—N$^{viii}$ | 117 (7) |
| I2$^{ii}$—Pb—I2$^{iv}$ | 90.000 (1) | N$^{vi}$—C—N$^{viii}$ | 55 (10) |
| I2$^{iii}$—Pb—I2$^{iv}$ | 90.000 (1) | C$^{vi}$—C—N | 29 (3) |
| I1—Pb—I2$^{v}$ | 89.87 (16) | N$^{vi}$—C—N | 101 (9) |
| I1$^{i}$—Pb—I2$^{v}$ | 90.13 (16) | N$^{viii}$—C—N | 88 (9) |
| I2$^{ii}$—Pb—I2$^{v}$ | 90.000 (1) | C$^{vi}$—C—N$^{vii}$ | 29 (3) |
| I2$^{iii}$—Pb—I2$^{v}$ | 90.000 (1) | N$^{vi}$—C—N$^{vii}$ | 88 (9) |
| I2$^{iv}$—Pb—I2$^{v}$ | 179.7 (3) | N$^{viii}$—C—N$^{vii}$ | 101 (9) |
| Pb—I1—Pb$^{ix}$ | 180.0 | N—C—N$^{vii}$ | 29 (6) |
| Pb$^{x}$—I2—Pb$^{xi}$ | 163.36 (10) | | |

Symmetry codes: (i) -*x*, *y*, *z*-1/2; (ii) *y*-1, -*x*, *z*; (iii) -*y*+1, *x*, *z*; (iv) *x*, *y*-1, *z*; (v) -*x*, -*y*+1, *z*; (vi) -*x*+1, -*y*, *z*; (vii) -*y*+1/2, -*x*+1/2, *z*; (viii) *y*+1/2, *x*-1/2, *z*; (ix) -*x*, *y*, *z*+1/2; (x) -*x*+1/2, *y*+1/2, *z*; (xi) *x*,

*y*+1, *z*.

## 4. Split site model refinement

Refined as a 2-component inversion twin with a twin fraction of

*Table S7: Crystal data*

| | |
|---|---|
| $CH_6I_3NPb$ | *Z* = 4 |
| $M_r$ = 619.96 | *F*(000) = 1040 |
| Tetragonal, *I*4*cm* | $D_x$ = 4.171 Mg $m^{-3}$ |
| *a* = 8.8438 (3) Å | μ = 26.39 $mm^{-1}$ |
| *c* = 12.6215 (5) Å | *T* = 293 K |
| *V* = 987.16 (8)  $Å^3$ | |

*Table S8: Data collection*

| | |
|---|---|
| 5668 measured reflections | $\theta_{max}$ = 34.5°, $\theta_{min}$ = 3.7° |
| 768 independent reflections | *h* = -12→12 |
| 674 reflections with $I > 2\sigma(I)$ | *k* = -12→12 |
| $R_{int}$ = 0.054 | *l* = -17→17 |

*Table S9: Refinement*

| | |
|---|---|
| Refinement on $F^2$ | H-atom parameters not defined |
| Least-squares matrix: full | $w = 1/[\sigma^2(F_o^2) + (0.0592P)^2 + 5.3133P]$ where $P = (F_o^2 + 2F_c^2)/3$ |
| $R[F^2 > 2\sigma(F^2)]$ = 0.035 | $(\Delta/\sigma)_{max}$ = 0.002 |
| $wR(F^2)$ = 0.113 | $\Delta\rangle_{max}$ = 2.03 e $Å^{-3}$ |
| *S* = 1.19 | $\Delta\rangle_{min}$ = -1.06 e $Å^{-3}$ |
| 768 reflections | Absolute structure:  Refined as an inversion twin. |
| 27 parameters | Absolute structure parameter: 0.49 (3) |
| 2 restraints | |

*Table S10: Fractional atomic coordinates and isotropic or equivalent isotropic displacement parameters ($Å^2$)*

| | *x* | *y* | *z* | $U_{iso}$*/$U_{eq}$ | Occ. (<1) |
|---|---|---|---|---|---|
| Pb | 0.000000 | 0.000000 | -0.0026 (5) | 0.0300 (2) | |
| I1 | 0.000000 | 0.000000 | 0.2473 (9) | 0.0814 (8) | |
| I2 | 0.2078 (4) | 0.7078 (4) | 0.000000 | 0.0549 (8) | 0.5616 (4) |
| I21 | 0.2442 (11) | 0.7442 (11) | -0.0206 (10) | 0.0549 (8) | 0.1829 (4) |
| I22 | 0.2145 (10) | 0.7145 (10) | 0.0353 (8) | 0.0549 (8) | 0.2555 (4) |
| N | 0.420 (6) | 0.046 (6) | 0.275 (4) | 0.044 (13)* | 0.25 |
| C | 0.540 (3) | -0.040 (3) | 0.221 (5) | 0.064 (14)* | 0.5 |

*Table S11: Atomic displacement parameters (Å$^2$)*

| | $U^{11}$ | $U^{22}$ | $U^{33}$ | $U^{12}$ | $U^{13}$ | $U^{23}$ |
|---|---|---|---|---|---|---|
| Pb | 0.0314 (3) | 0.0314 (3) | 0.0273 (4) | 0.000 | 0.000 | 0.000 |
| I1 | 0.1128 (13) | 0.1128 (13) | 0.0187 (7) | 0.000 | 0.000 | 0.000 |
| I2 | 0.0570 (7) | 0.0570 (7) | 0.051 (2) | 0.0348 (8) | 0.010 (2) | 0.010 (2) |
| I21 | 0.0570 (7) | 0.0570 (7) | 0.051 (2) | 0.0348 (8) | 0.010 (2) | 0.010 (2) |
| I22 | 0.0570 (7) | 0.0570 (7) | 0.051 (2) | 0.0348 (8) | 0.010 (2) | 0.010 (2) |

*Table S12: Geometric parameters (Å, º)*

| | | | |
|---|---|---|---|
| Pb—I21$^{i}$ | 3.1359 (11) | Pb—I2$^{ii}$ | 3.1711 (9) |
| Pb—I21$^{ii}$ | 3.1359 (11) | Pb—I2$^{iii}$ | 3.1711 (9) |
| Pb—I21$^{iii}$ | 3.1359 (11) | Pb—I2$^{iv}$ | 3.1711 (9) |
| Pb—I21$^{iv}$ | 3.1359 (11) | Pb—I22$^{ii}$ | 3.194 (2) |
| Pb—I1 | 3.154 (7) | Pb—I22$^{i}$ | 3.194 (2) |
| Pb—I1$^{v}$ | 3.157 (7) | N—N$^{vi}$ | 0.43 (12) |
| Pb—I2$^{i}$ | 3.1711 (9) | | |
| | | | |
| I21$^{i}$—Pb—I21$^{iii}$ | 89.70 (4) | I21$^{ii}$—Pb—I2$^{iv}$ | 98.3 (3) |
| I21$^{ii}$—Pb—I21$^{iii}$ | 89.70 (4) | I21$^{iii}$—Pb—I2$^{iv}$ | 171.0 (3) |
| I21$^{i}$—Pb—I21$^{iv}$ | 89.70 (4) | I21$^{iv}$—Pb—I2$^{iv}$ | 9.5 (2) |
| I21$^{ii}$—Pb—I21$^{iv}$ | 89.70 (4) | I1—Pb—I2$^{iv}$ | 89.42 (11) |
| I21$^{iii}$—Pb—I21$^{iv}$ | 171.7 (5) | I1$^{v}$—Pb—I2$^{iv}$ | 90.58 (11) |
| I21$^{i}$—Pb—I1 | 94.2 (3) | I2$^{i}$—Pb—I2$^{iv}$ | 89.994 (2) |
| I21$^{ii}$—Pb—I1 | 94.2 (3) | I2$^{ii}$—Pb—I2$^{iv}$ | 89.994 (2) |
| I21$^{iii}$—Pb—I1 | 94.2 (3) | I2$^{iii}$—Pb—I2$^{iv}$ | 178.8 (2) |
| I21$^{iv}$—Pb—I1 | 94.2 (3) | I1—Pb—I22$^{ii}$ | 81.40 (19) |
| I1—Pb—I1$^{v}$ | 180.0 | I1$^{v}$—Pb—I22$^{ii}$ | 98.60 (19) |
| I1—Pb—I2$^{i}$ | 89.42 (11) | I2$^{i}$—Pb—I22$^{ii}$ | 170.7 (3) |
| I1$^{v}$—Pb—I2$^{i}$ | 90.58 (11) | I2$^{ii}$—Pb—I22$^{ii}$ | 8.15 (19) |
| I1—Pb—I2$^{ii}$ | 89.42 (11) | I2$^{iii}$—Pb—I22$^{ii}$ | 88.4 (3) |
| I1$^{v}$—Pb—I2$^{ii}$ | 90.58 (11) | I2$^{iv}$—Pb—I22$^{ii}$ | 91.4 (3) |
| I2$^{i}$—Pb—I2$^{ii}$ | 178.8 (2) | I1—Pb—I22$^{i}$ | 81.40 (19) |
| I21$^{i}$—Pb—I2$^{iii}$ | 98.3 (3) | I1$^{v}$—Pb—I22$^{i}$ | 98.60 (19) |
| $I21^{ii}$—Pb—$I2^{iii}$ | 81.8 (3) | $I2^{i}$—Pb—$I22^{i}$ | 8.15 (19) |
| $I21^{iii}$—Pb—$I2^{iii}$ | 9.5 (2) | $I2^{ii}$—Pb—$I22^{i}$ | 170.7 (3) |
| $I21^{iv}$—Pb—$I2^{iii}$ | 171.0 (3) | $I2^{iii}$—Pb—$I22^{i}$ | 91.4 (3) |
| I1—Pb—$I2^{iii}$ | 89.42 (11) | $I2^{iv}$—Pb—$I22^{i}$ | 88.4 (3) |
| $I1^{v}$—Pb—$I2^{iii}$ | 90.58 (11) | Pb—I1—$Pb^{vii}$ | 180.0 |
| $I2^{i}$—Pb—$I2^{iii}$ | 89.994 (2) | $Pb^{viii}$—I2—$Pb^{ix}$ | 160.82 (19) |
| $I2^{ii}$—Pb—$I2^{iii}$ | 89.994 (2) | $Pb^{ix}$—I21—$Pb^{viii}$ | 171.3 (5) |
| $I21^{i}$—Pb—$I2^{iv}$ | 81.8 (3) | $Pb^{viii}$—I22—$Pb^{ix}$ | 156.4 (4) |

Symmetry codes: (i) *y*-1, -*x*, *z*; (ii) -*y*+1, *x*, *z*; (iii) -*x*, -*y*+1, *z*; (iv) *x*, *y*-1, *z*; (v) -*x*, *y*, *z*-1/2; (vi) -*y*+1/2, -*x*+1/2, *z*; (vii) -*x*, *y*, *z*+1/2; (viii) -*x*+1/2, *y*+1/2, *z*; (ix) *x*, *y*+1, *z*.